\documentclass{iau}
\usepackage{natbib}

\title[Sgr A* accretion flow from GR dynamical and radiative modeling] 
{Constraining the Accretion Flow in Sgr A* by General Relativistic Dynamical and Polarized Radiative Modeling}

\author[Roman V. Shcherbakov et al.]   
{Roman V. Shcherbakov $^{1,2}$, Robert F. Penna $^3$ \and Jonathan C. McKinney $^4$}

\affiliation{$^1$ Department of Astronomy, University of Maryland, College Park, MD 20742-2421, USA \\ email: {\tt roman@astro.umd.edu} \\[\affilskip]
$^2$Hubble Fellow\\[\affilskip]
$^3$Harvard-Smithsonian Center for Astrophysics, 60 Garden Street, Cambridge, MA 02138, USA \\[\affilskip]
$^4$Physics Department, University of Maryland, College Park, MD 20742-4111, USA}

\pubyear{2012}
\volume{290}  
\jname{Feeding compact objects: Accretion on all scales}
\editors{C.M. Zhang, T. Belloni, M. M\'endez \& S.N. Zhang, eds.}

\begin{document}
\maketitle
\begin{abstract}
We briefly summarize the method of simulating Sgr A* polarized sub-mm spectra from the accretion flow and fitting the observed spectrum.
The dynamical flow model is based on three-dimensional general relativistic magneto hydrodynamic simulations.
Fully self-consistent radiative transfer of polarized cyclo-synchrotron emission is performed.
We compile a mean sub-mm spectrum of Sgr A* and fit it with the mean simulated spectra. We estimate the ranges of inclination angle $\theta=42^\circ-75^\circ$, mass accretion rate $\dot{M}=(1.4-7.0)\times10^{-8}M_\odot{\rm year}^{-1}$, and  electron temperature $T_e=(3-4)\times10^{10}$K at $6M$.
We discuss multiple caveats in dynamical modeling, which must be resolved to make further progress.
\end{abstract}

\firstsection

\section{Estimating Sgr A* BH and accretion flow parameters}
Sagittarius A* (Sgr A*) is a supermassive black hole (BH) in the center of the Milky Way galaxy.
Its accretion flow is underluminous and radiatively inefficient \citep{Yuan:2003sg,Sharma_heating:2007} for the radial profiles of electron temperature $T_e$
consistent with observations. Thus the gas infall can be modeled with three-dimensional (3D) general relativistic (GR) magneto hydrodynamic (MHD) simulations without cooling. Such simulations and the associated radiative transfer were recently performed by \cite{Moscibrodzka:2009,Dexter:2010lk,Hilburn:2010dh} and by our group
\citep{Shcherbakov:2012appl}, while some groups resorted to (semi-)analytic accretion flow models \citep{Huang:2009ma,Broderick:2011sg}.
All researchers aimed at fitting combinations of radio, (polarized) sub-mm, IR, and X-ray spectra and emitting region size.

Our group adopts the following approach in fitting Sgr A* observations and estimating the accretion flow and the BH parameters.
We compile the mean Sgr A* polarized sub-mm spectrum from observations in the past $\sim10$~years. We perform a set of 3D GRMHD simulations without cooling of accretion flow onto the BH for several spin values. We apply the original GR polarized radiative transfer code, which follows the prescription in \cite{Shcherbakov:2011inter} and employs Faraday rotation and conversion coefficients from \cite{Shcherbakov:2008fa}. The radiative transfer is fully self-consistent. The mean simulated spectrum is found as a mean of spectra at different simulation times.
We explore the parameter space of spin $a_*$, inclination angle $\theta$, mass accretion rate $\dot{M}$, and the ratio of proton to electron temperature $T_p/T_e$ at radius $6M$. For each spin we find a model, which provides the best agreement between the mean simulated and the mean observed spectra.
We test our numerical radiative transfer code and the results for robustness. Our best bet model has $a_*=0.5$, $\theta=75^\circ$, $\dot{M}=4.6\times10^{-8}M_\odot{\rm year}^{-1}$, $T_e=3.1\times10^{10}$~K at $6M$, and spin position angle ${\rm PA}=115^\circ$. Model parameters are in the ranges $\theta=42^\circ-75^\circ$, $\dot{M}=(1.4-7.0)\times10^{-8}M_\odot{\rm year}^{-1}$, and $T_e=(3-4)\times10^{10}$K at $6M$ across all spins. The sub-mm circular polarization is mainly produced by Faraday conversion as modified by Faraday rotation. The emission region size at $230$~GHz is consistent with the VLBI size \citep{Doeleman:2008af}. The details of the modeling can be found in our main paper \citep{Shcherbakov:2012appl}.
\section{Caveats}
Present-day dynamical modeling of tenuous plasma falling onto the BH inevitably relies on heavy approximations.
Our numerical resolution in MHD simulations might not be sufficient to achieve convergence in turbulence dissipation rate or magnetic field structure \cite{Hawley:2011po}.
The dynamic range and duration of the simulation are not large enough to obtain reliable radial profiles of quantities and the long-term behavior of the accretion flow.
MHD approximation itself might not hold for collisionless plasma.
 Non-trivial plasma effects include non-thermal distribution of particles inferred for both the flared \cite{Dodds-Eden:2009} and the quiescent \cite{Yuan:2003sg} states. Another important effect is thermal heat conduction. It was found to dramatically alter the radial density profile and other quantities \citep{Johnson:2007qw,Shcherbakov:2010cond}.
In view of the aforementioned difficulties the results of our modeling should be viewed as estimates with unknown systematics.
We have developed the framework of testing various dynamical models of Sgr A* and other low-luminosity active galactic nuclei.
Whenever the more self-consistent dynamical models become available, it will be possible to promptly apply them to fit Sgr A* accretion flow.

\acknowledgements The work is partially supported by NASA Hubble Fellowship grant HST-HF-51298.01.

\end{document}